\documentclass{ethpaper}
\usepackage{graphicx}
\usepackage{here}
\usepackage{amssymb}
\usepackage{amsmath}
\usepackage{subfigure}
\usepackage{lineno}
\usepackage{wasysym}

\begin{document}
\begin{titlepage}
\ethnote{}
\title{Performance studies of scintillating ceramic samples exposed to ionizing radiation}
\begin{Authlist}
G.~Dissertori, D.~Luckey, \underline{F.~Nessi-Tedaldi}, F.~Pauss, R.~Wallny\\Institute for Particle Physics, ETH Zurich, 8093 Zurich, Switzerland
\end{Authlist}
\maketitle
\begin{abstract}
Scintillating ceramics are a promising, new development for various applications in science and industry. 
Their application in calorimetry for particle physics experiments is expected to involve an exposure to high levels of ionizing radiation. In this paper, changes in performance have been measured for scintillating ceramic samples of different composition after exposure to penetrating ionizing radiation up to a dose of 38 kGy.
\end{abstract}
\vspace{7cm}
\conference{\em To be published in the 2012 IEEE Nuclear Science Symposium Conference Record}
\end{titlepage}
\section{Introduction}

Sintered, transparent materials ("ceramics") are being developed as a competitive alternative to crystals for use as scintillators in various fields of application. While homeland security and commercial applications are the primary fields of envisaged applications, the scientific community is showing an increasing interest in their potential use as heavy scintillators in calorimetry. Where large detectors are needed,  as for particle physics experiments, scintillating ceramics might offer in fact an affordable alternative to inorganic crystals. 

The environment in which a scintillating material will be used can place stringent requirements on performance, 
in particular where high levels of ionizing radiation are expected, as it will be the case for the High-Luminosity Large Hadron Collider (HL-LHC) at CERN.
For crystals, the achievement of the required radiation resistance is known to depend on the stoichiometric composition and on the suppression of defects at the ppm level, which in some cases has required many years of R\&D~\cite{r-RADQ,r-EA,r-ANN,r-LQB,r-MAO}. For ceramic scintillators, this is a yet unexplored territory, and no material optimization has been performed so far to meet such requirements.

It was thus mandatory, as a first step in the exploration of the ceramics' potential, to test the effects of ionizing radiation on available samples. In this paper, we will present first results for the effect of ionizing radiation on scintillating ceramic samples after irradiation with $^{60}{\mbox{Co}}$ $\gamma$ up to a dose of 38 kGy at a dose rate of 0.58 kGy/h, the latter being representative for the expected running conditions in the most exposed regions of an electromagnetic calorimeter at the HL-LHC. The studied samples --- of different composition and origin --- have not been previously optimized for radiation hardness, thus the presented results provide conservative limits on the achievable performance.

\section{Characteristics of scintillating ceramics}
\label{s-CHAR}
The production of transparent scintillating ceramics 
involves sintering nanocrystals into fully dense, transparent polycrystalline solids. Transparent ceramics production offers an alternative to crystal growth, and is expected to be more economical, especially for high-melting point materials, since efficient sintering occurs at temperatures below the melting point. 
Sintering involves a faster production cycle compared to crystal growth in high-temperature furnaces, and samples can be formed to shape. The resulting samples can be machined, are robust, and unlike crystals, exhibit no cleavage planes. In case an activator for scintillation is added, its distribution is inherently uniform across the sample, while segregation is always a problem when growing large crystals.
The present, early stage of the  production method development, which still limits the size of available samples, is one disadvantage, along with the fact that sintering is possible only where the crystalline structure is cubic.

\section{Possible applications}
\label{s-POSS}
We have studied in this work ceramics at their present stage of development, to establish whether their would be suitable 
scintillators for a calorimetry upgrade aimed at running at the HL-LHC.
\begin{figure}[b]
\centering
\includegraphics[width=3.2in]{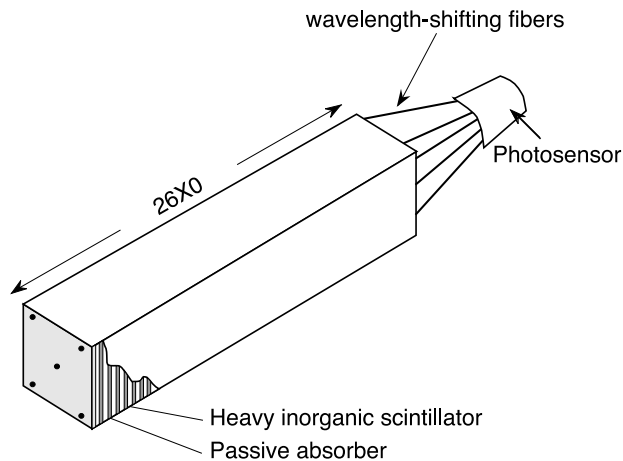}
\caption{Schematic drawing of a Shashlik module}
\label{f-SHAS}
\end{figure}
There, the main issue will be ensuring an adequate performance under exposure to ionizing radiation levels up to
0.15 kGy/h and 1 MGy in total~\cite{r-UTP}, and to energetic hadron fluences up to $3\times 10^{14}\;\mathrm{particles/cm}^2$.
While the present stage of ceramics production may not allow, for several years, envisaging the construction of a homogeneous calorimeter, scintillating ceramic plates a few millimeters in thickness already reach the suitable dimensions for a sampling
calorimeter. A sampling calorimeter with alternating layers of passive absorber (e.g. Pb or W) and active plates of a heavy scintillating ceramics provided with a "Shashlik"-type readout~\cite{r-SHAS}, as shown in Fig.~\ref{f-SHAS}, could be built to have  compact dimensions, comparable to a currently running homogeneous calorimeter such as the CMS ECAL~\cite{r-ECALTDR}.
\section{Ionizing radiation test procedure}
We have studied the change of light transmission for the available samples due to exposure to penetrating ionizing radiation. 
As it is well known, ionizing radiation causes color centers to materials, that tend to absorb light. The damage, in a given material,
tends to reach an equilibrium between spontaneous recovery at room temperature and damage, whose amplitude depends on dose rate. We have exposed the samples to a ionizing dose rate $\phi_{\gamma} = 0.58\;\mathrm{kGy/h}$ for a total integrated dose
$\Phi_{\gamma} = 38\;\mathrm{kGy}$. After irradiation, all the samples were kept in the dark, at room temperature, except for the few moments needed for insertion and removal in the spectrophotometer. In the test, the dose rate used is somewhat higher than expected for HL-LHC running, it is
however representative for the ionizing dose rate accompanying hadron-irradiations that we envisage performing at a later
stage. To quantify the damage, we have measured the Light Transmission (LT) as a function of wavelength before and after irradiation, and we have determined the induced absorption coefficient $\mu_{IND}(\lambda)$ as a function of light wavelength $\lambda$, as explained in~\cite{r-LTNIM}, as:
\begin{equation}
\mu_{IND}(\lambda) = \frac{1}{\ell}\times \ln \frac{LT_0 (\lambda)}{LT (\lambda)}
\label{muDEF}
\end{equation}
where $LT_0\; (LT)$ is the Light Transmission value measured before (after) irradiation through the thickness $\ell$ of the crystal.

\section{Scintillating ceramic samples and results}
We have tested a few scintillating ceramics samples, that have been developed for purposes other than calorimetry in 
high-energy hadron collisions. Thus, we did not expect them {\em a priori} to be optimized in their radiation hardness.
All the studied samples contain Cerium as the activator for scintillation, and emit a broad spectrum of scintillation light, which is peaked around 520 nm.
\subsection{Lutetium-Yttrium garnets}
\begin{table}[b]
\caption{Lutetium-Yttrium ceramic garnets.\label{t-T1}}
\centering
\begin{tabular}{|l|l|c|c|c|c|}
\hline
Label & Composition & $\diameter$ & $\ell$  & $\rho$       & $X_0$\\
           &                         & [mm] & [mm] & [g/cm$^3$]      & [cm]     \\
\hline
\hline
cYAG1 &     ${\mathrm{Y_3 Al_5 O_{12}}}$                      &$22.7$&$5.1$ & $4.3$ & $3.7$ \\
cYAG2 &     ${\mathrm{Y_3 Al_5 O_{12}}}$                      &$14.0$&$2.1$ & $4.3$ & $3.7$ \\
cLuYAG& ${\mathrm{(Y,Lu)_3Al_5O_{12}}}$ &$23.4$&$1.3$ & $5.5$ & $2.1$ \\
cLuAG &  ${\mathrm{Lu_3 Al_5 O_{12}}}$                      &$14.0$&$1.0$ & $6.6$ & $1.4$ \\
\hline
\end{tabular}
\end{table}
\begin{figure}[!b]
\centering
\includegraphics[width=5in]{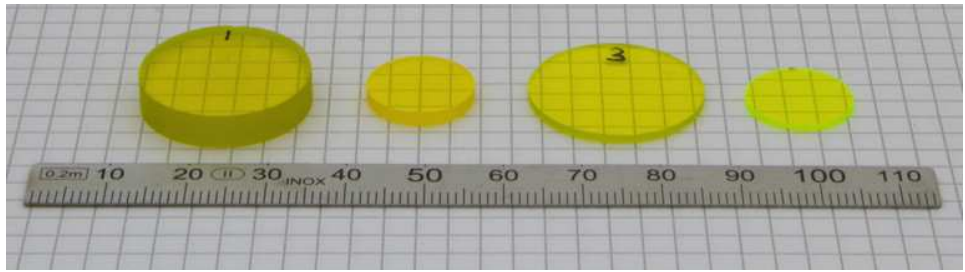}
\caption{Lutetium-Yttrium ceramic garnet samples.}
\label{f-allLuYAG}
\end{figure}
\begin{figure}[!t]
\begin{center}\footnotesize
\begin{tabular}[ch]{cc}
\subfigure[for $200 < \lambda < 800$ nm.]
{\mbox{\includegraphics[width=1.5in,trim=138mm 53mm 5mm 5mm , clip]{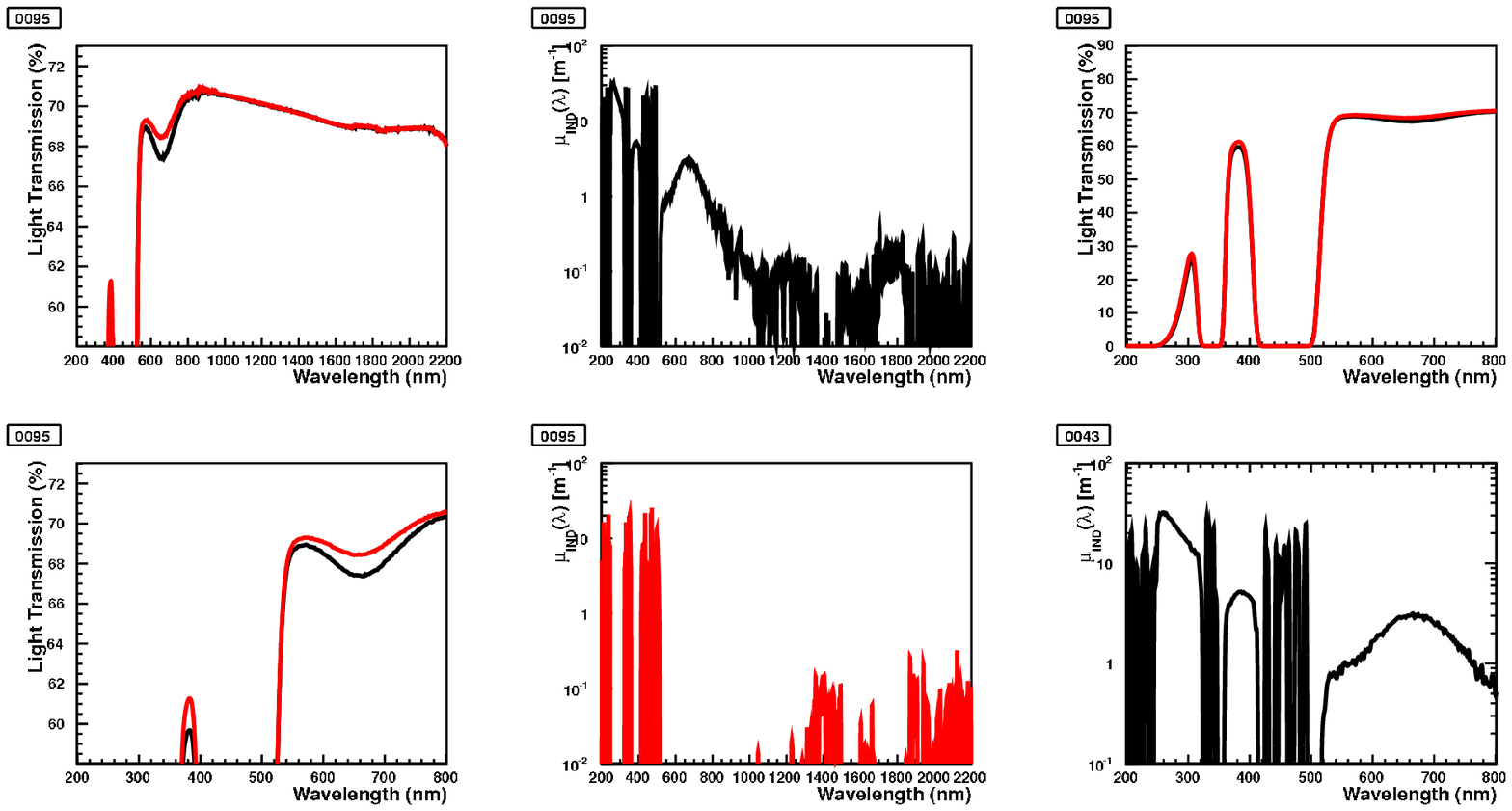}
}}&
\subfigure[zoomed detail]
{\mbox{\includegraphics[width=1.5in, trim=5mm 0mm 138mm 58mm , clip]{YSO0095mu.eps}
}}
\end{tabular}
\caption{Light Transmission before (in red) and after (in black) irradiation, for sample cYAG1.\label{f-YAG1}}
\end{center}
\end{figure}
\begin{figure}[!t]
\begin{center}\footnotesize
\begin{tabular}[ch]{cc}
\subfigure[for $200 < \lambda < 800$ nm.]
{\mbox{\includegraphics[width=1.5in,trim=138mm 53mm 5mm 5mm , clip]{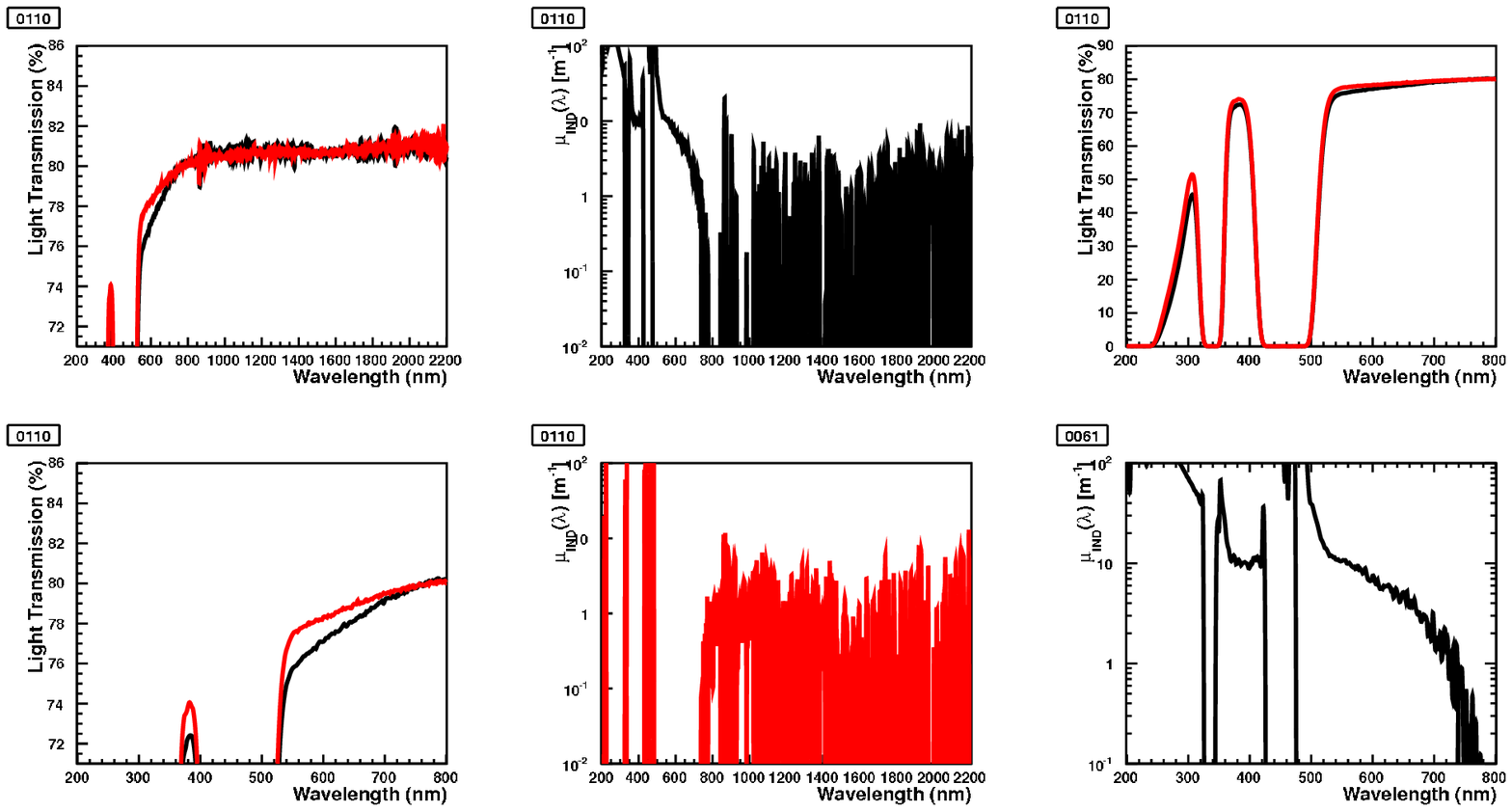}
}}&
\subfigure[zoomed detail]
{\mbox{\includegraphics[width=1.5in, trim=5mm 0mm 138mm 58mm , clip]{YSO0110mu.eps}
}}
\end{tabular}
\caption{Light Transmission before (in red) and after (in black) irradiation, for sample cYAG2.\label{f-YAG2}}
\end{center}
\end{figure}
\begin{figure}[!t]
\begin{center}\footnotesize
\begin{tabular}[ch]{cc}
\subfigure[for $200 < \lambda < 800$ nm.]
{\mbox{\includegraphics[width=1.5in,trim=138mm 53mm 5mm 5mm , clip]{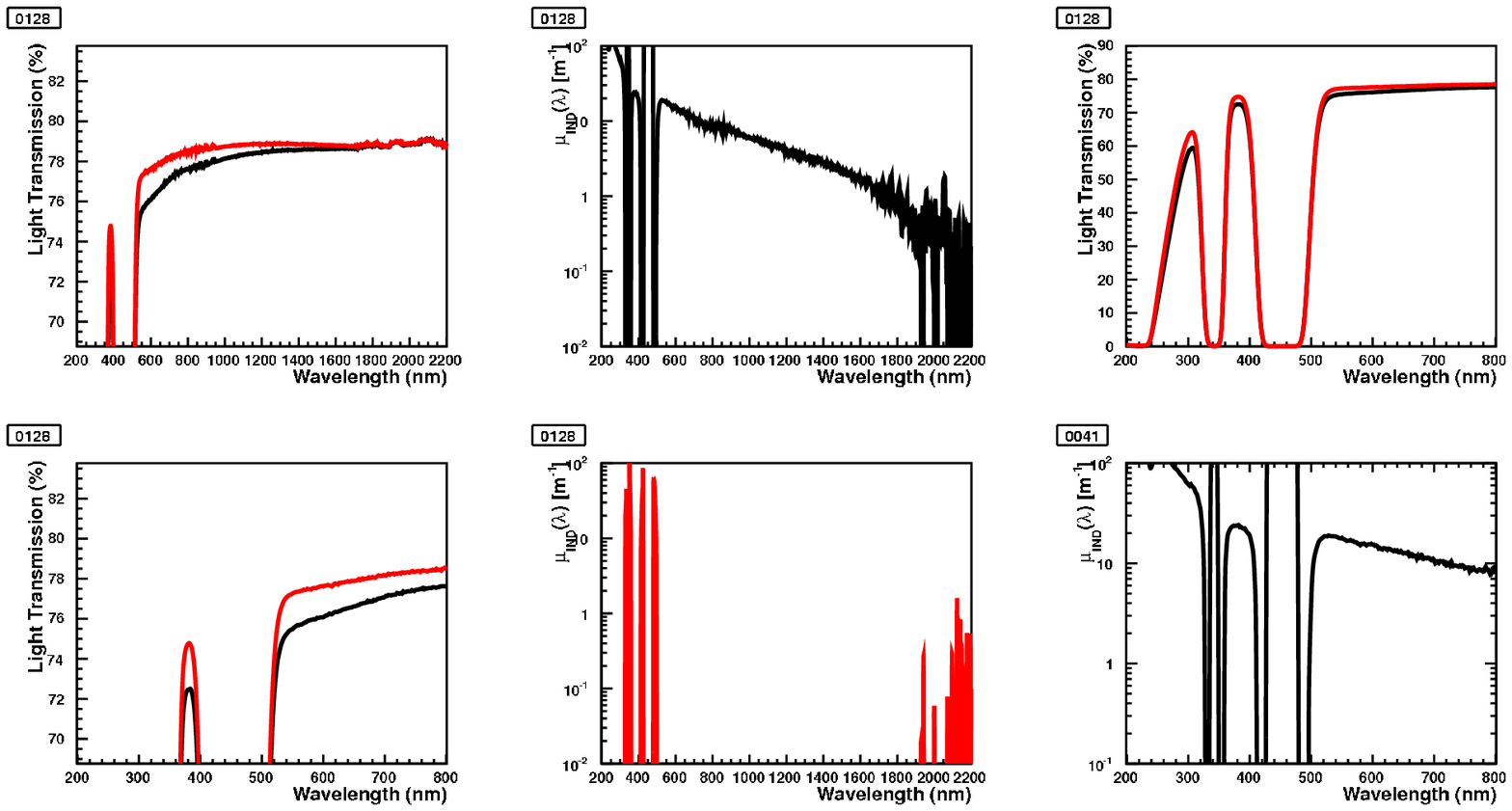}
}}&
\subfigure[zoomed detail]
{\mbox{\includegraphics[width=1.5in, trim=5mm 0mm 138mm 58mm , clip]{YSO0128mu.eps}
}}
\end{tabular}
\caption{Light Transmission before (in red) and after (in black) irradiation, for sample cLuYAG.\label{f-LuYAG}}
\end{center}
\end{figure}
\begin{figure}[!t]
\begin{center}\footnotesize
\begin{tabular}[ch]{cc}
\subfigure[for $200 < \lambda < 800$ nm.]
{\mbox{\includegraphics[width=1.5in,trim=138mm 53mm 5mm 5mm , clip]{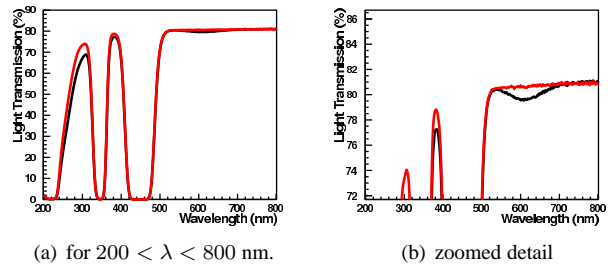}
}}&
\subfigure[zoomed detail]
{\mbox{\includegraphics[width=1.5in, trim=5mm 0mm 138mm 58mm , clip]{YSO0108mu.eps}
}}
\end{tabular}
\caption{Light Transmission before (in red) and after (in black) irradiation, for sample LuAG.\label{f-LuAG}}
\end{center}
\end{figure}
The Lutetium-Yttrium garnets we have tested, have been kindly provided by the Shanghai Institute of Ceramics~\cite{r-SIC}, where they are being developed mainly for applications as LED components. For this reason, they exhibit important fluorescence levels, that would need to be minimized for HL-LHC applications. They are visible in Fig.~\ref{f-allLuYAG} and their characteristics are listed, along with their composition, in Table~\ref{t-T1}, where $\rho$ is the density, $\ell$ the thickness and $X_0$ the calculated radiation length.
The Light Transmissions before and after
irradiation for the various samples are shown in Figs.~\ref{f-YAG1} to \ref{f-LuAG}. For all the samples, a broad color center is visible, 
that however affects the emission region in a modest way. For sample cYAG1, the induced absorption $\mu_{IND}$ does not exceed $1\;{\mbox m}^{-1}$ over the entire range of emission wavelengths, between 520 and 560 nm. For samples cYAG2 and cLuYAG, the value stays below $10\;{\mbox m}^{-1}$ and it remains smaller than $4\;{\mbox m}^{-1}$ in cLuAG.
No recovery of damage has been observed over several months.

\subsection{GYGAG garnet}
\begin{figure}[h]
\centering
\includegraphics[width=2in]{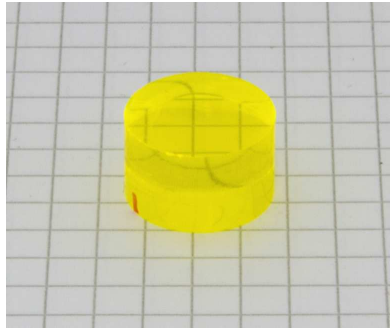}
\caption{Ceramic GYGAG sample from LLNL. }
\label{f-photoGYG}
\end{figure}

The GYGAG garnet we have tested, has been kindly made available to us by the Lawrence Livermore National Laboratory~\cite{r-LLNL}.\begin{table}[!h]
\caption{GYGAG ceramic garnet\label{t-T2}}
\centering
\begin{tabular}{|l|l|c|c|c|c|c|}
\hline
Label & Composition & $\diameter$ & $\ell$  & $\rho$&  $X_0$\\
           &                         & [mm] & [mm] & [g/cm$^3$]&  [cm]     \\
\hline
\hline
cGYGAG&${\mathrm{(Gd,Y)_3(Ga,Al)_5O_{12}}}$&$13$&$8$ & $4.1$ & $2$ \\
\hline
\end{tabular}
\end{table}
 Its characteristics are found in Table~\ref{t-T2} and the sample is visible in Fig~\ref{f-photoGYG}. Concerning this sample, whose composition is ${\mathrm{(Gd,Y)_3(Ga,Al)_5O_{12}}}$,
it should be noted that it contains the $^{157}{\mathrm{Gd}}$ isotope in its natural abundance. This isotope is known for having
an extraordinarily high capture cross-section for thermal neutrons, of 290'000 barn, that makes it unsuitable for
calorimetry at HL-LHC, where high fluxes of thermal neutrons are expected during running.

\begin{figure}[!h]
\begin{center}\footnotesize
\begin{tabular}[ch]{cc}
\subfigure[for $200 < \lambda < 800$ nm.]
{\mbox{\includegraphics[width=1.5in,trim=138mm 53mm 5mm 5mm , clip]{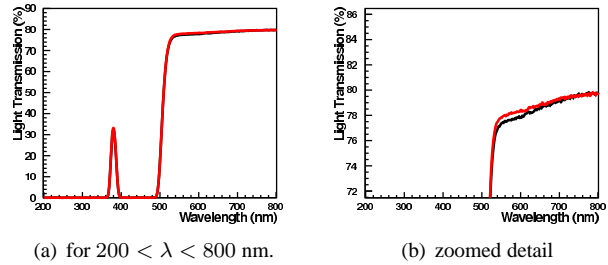}
}}&
\subfigure[zoomed detail]
{\mbox{\includegraphics[width=1.5in, trim=5mm 0mm 138mm 58mm , clip]{YSO0113mu.eps}
}}
\end{tabular}
\caption{Light Transmission before (in red) and after (in black) irradiation, for sample cGYGAG.\label{f-GYG}}
\end{center}
\end{figure}
The Light Transmission before and after
irradiation for cGYGAG is shown in Fig.~\ref{f-GYG}. Here as well, a broad color center is visible, of quite small amplitude.
The induced absorption $\mu_{IND}$ does not exceed $1\;{\mbox m}^{-1}$ over the entire range of emission wavelengths, between 520 and 580 nm.  For this sample as well, no recovery of damage has been observed over several months.

\section*{Conclusions}
Several transparent scintillating ceramic samples have been exposed to important levels of ionizing radiation, and the changes in light Transmission have been measured. The damage, in terms of induced absorption length, is modest, compared to the dimension of the samples, were scintillation light to be transported through them.
The magnitude of ionizing radiation damage at the wavelengths of scintillation emission might be adequate for their use in a sampling calorimeter at the HL-LHC, where light would have to travel $\sim 2$ cm inside the scintillator before being collected.

\section*{Acknowledgements}

We are indebted to Prof. X.~Q.~ Feng and his colleagues from SIC, as well as to Prof. N. Cherepy from LLNL for providing the scintillating ceramic samples for our tests. We are grateful to V. Boutellier from the Paul Scherrer Institute in Villigen, Switzerland, for the irradiation of the samples. This work was partially supported by the Swiss National Science Foundation.

\end{document}